# Attenuation of superconductivity in manganite/cuprate heterostructures by epitaxially-induced CuO intergrowths

H. Zhang,[1] N. Gauquelin,[2] G. A. Botton,[2] and J. Y. T. Wei[1,3]

[1] *Department of Physics, University of Toronto, Toronto ON M5S1A7 Canada*
[2] *Brockhouse Institute for Materials Research, McMaster University, Hamilton ON L8S4M1 Canada*
[3] *Canadian Institute for Advanced Research, Toronto ON M5G1Z8 Canada*

We examine the effect of CuO intergrowths on the superconductivity in epitaxial $La_{2/3}Ca_{1/3}MnO_3$/ $YBa_2Cu_3O_{7-\delta}$ (LCMO/YBCO) thin-film heterostructures. Scanning transmission electron microscopy on bilayer LCMO/YBCO thin films revealed double CuO-chain intergrowths which form regions with the 247 lattice structure in the YBCO layer. These nanoscale 247 regions do not appear in x-ray diffraction, but can physically account for the reduced critical temperature ($T_c$) of bilayer thin films relative to unilayer films with the same YBCO thickness, at least down to $\sim 25$ nm. We attribute the CuO intergrowths to the bilayer heteroepitaxial mismatch and the $T_c$ reduction to the generally lower $T_c$ seen in bulk 247 samples. These epitaxially-induced CuO intergrowths provide a microstructural mechanism for the attenuation of superconductivity in LCMO/YBCO heterostructures.

PACS numbers: 74.78.Fk, 74.72.-h, 75.47.Lx, 68.37.Og

The lattice compatibility among transition-metal oxides enables them to be epitaxially combined in thin-film form.[1] In recent years, there have been numerous studies of heteroepitaxial thin films comprising the ferromagnetic (F) manganites and superconducting (S) cuprates, probing novel effects of the F/S interplay ranging from spin injection to proximity coupling.[2–7] An observation of particular interest is the dependence of the superconducting critical temperature ($T_c$) on the $c$-axis layer thicknesses in multilayer $La_{2/3}Ca_{1/3}MnO_3/YBa_2Cu_3O_{7-\delta}$ (LCMO/YBCO) films.[8,9] The length scale of this dependence indicates an extremely long-ranged F/S proximity effect which is of both technological and theoretical interest,[10–12] although direct and microscopic evidence for this effect is still elusive.[13,14] Other interfacial mechanisms, such as charge transfer,[15–17] orbital reconstruction,[18] spin diffusion[9] and induced magnetic modulation,[19,20] are also believed to affect the superconductivity in LCMO/YBCO heterostructures.

A crucial aspect of LCMO/YBCO heterostructures that has not been well studied is the microscopic stoichiometry of the YBCO layer. The Y-Ba-Cu-O compounds are exceptional among the cuprates in having CuO chains, the number of which per unit cell allows the cation stoichiometry to vary between the so-called 123, 124, and 247 phases, which we denote as YBCO-123, YBCO-124, and YBCO-247. These phases have different optimal $T_c$, with bulk YBCO-247 showing generally lower $T_c$ than either YBCO-124 or fully-oxygenated YBCO-123.[21–27] As shown in Figure 1(a), the 123 and 124 phases have single and double CuO chains, respectively, while the 247 phase consists of alternating 123 and 124 blocks. Local stoichiometric variations have often been seen in nominally-123 YBCO samples, typically as intergrowths of extra CuO chains.[28–30] For sufficiently thin YBCO layers, such nanoscale intergrowths could constitute significant phase inhomogeneity in the paths of conduction, thus affecting the resistively determined $T_c$.

To elucidate the effect of CuO intergrowths on the superconductivity in LCMO/YBCO heterostructures, we carried out a microstructural study of bilayer LCMO/YBCO and unilayer YBCO thin films, in relation to their resistively measured $T_c$. Scanning transmission electron microscopy (STEM) on the bilayer films revealed double CuO-chain intergrowths which form YBCO-247 regions in the YBCO layer. These YBCO-247 regions do not show up in x-ray diffraction (XRD), but can physically account for the reduced $T_c$ measured in bilayer LCMO/YBCO films relative to unilayer YBCO films of the same YBCO thickness. We attribute the CuO intergrowths to the bilayer heteroepitaxial mismatch and the $T_c$ reduction to the generally lower $T_c$ seen in bulk 247 samples.

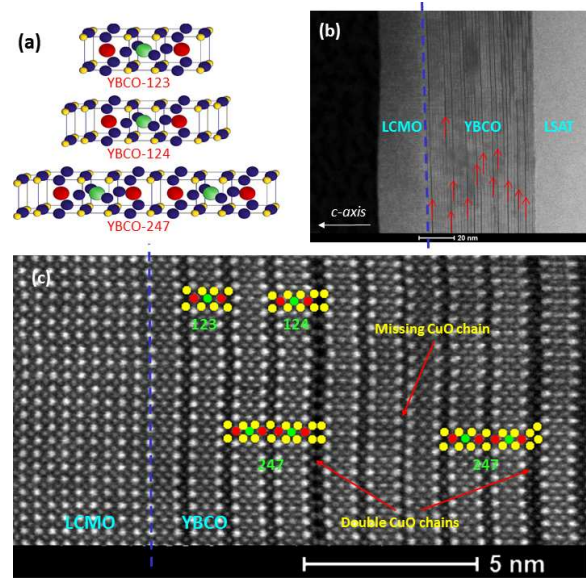

FIG. 1. HAADF-STEM images of a 25 nm/50 nm bilayer LCMO/YBCO film grown on (001)-oriented LSAT substrate. A low-resolution image is shown in panel (b), demonstrating uniform heteroepitaxy and layer thickness, with the LCMO/YBCO interface marked by blue dotted line. A high-resolution image near the LCMO/YBCO interface is shown in panel (c), revealing intergrowths of double CuO chains which form nanoscale 247 regions. These double chains are also visible in panel (b), as indicated by arrows. The lattice structures of YBCO-123, YBCO-124 and YBCO-247 phases are shown in panel (a), with the Cu, Y, and Ba atoms color-labeled as yellow, green, and red, respectively.

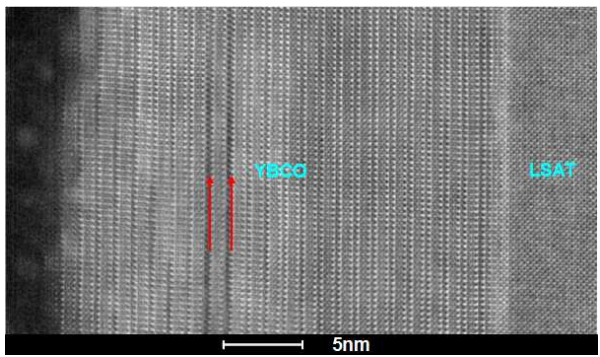

FIG. 2. STEM image of a 25 nm unilayer YBCO films grown on LSAT substrate, showing much less intergrowth of double-CuO chains (indicated by arrows) than the bilayer LCMO/YBCO film displayed in Fig. 1.

The bilayer LCMO/YBCO and unilayer YBCO films used in our study were epitaxially grown on (001)-oriented $(La,Sr)(Al,Ta)O_3$ (LSAT) substrates using pulsed laser-ablated deposition (PLD) with similar growth parameters as described in Ref. 13. The LCMO layer was 25 nm thick, while the YBCO layer was either 25 nm or 50 nm thick. LSAT was chosen for its close lattice matching with YBCO. Targets of LCMO and YBCO, each having $\sim 90\%$ material density and $> 99.9\%$ chemical purity, were used. The microstructure of the films was characterized using a FEI Titan 80-300 microscope fitted with a high-brightness field emission gun and CEOS aberration correctors for both condenser and objective lens aberrations. The microscope was operated at 200 keV, with the capability for elemental analysis using atomic-scale Electron Energy Loss Spectroscopy (EELS). XRD was also carried out on the films using the $\theta - 2\theta$ method with Cu $K_\alpha$ radiation from a Philips PW2273/20 X-ray tube. Finally, electrical resistance of the films was measured vs. temperature using standard ac lock-in technique in the four-contact configuration.

Figure 1 shows High-Angle Annular Dark-Field (HAADF) STEM images taken over the cross section of a 25 nm/50 nm bilayer LCMO/YBCO film grown on LSAT substrate. Fig. 1(b) shows a low-resolution image, demonstrating uniform heteroepitaxy and layer thickness. Fig. 1(c) shows a high-resolution image near the LCMO/YBCO interface; the color labels indicate different elements as identified from the HAADF-STEM images in which the contrast is sensitive to the atomic number (the higher atomic number the brighter the atomic column appears) and by atomic-scale EELS. Details of the elemental mapping by EELS will be published elsewhere.[31] This image shows the LCMO/YBCO interface consists of Mn atoms joining Ba atoms at CuO chain sites. This type of LCMO/YBCO interface is commonly reported for LCMO/YBCO heterostructures grown by either PLD or sputtering.[32–34] To the right of this interface, there are three defect-free unit cells of YBCO-123 characterized by single CuO chains between Ba atoms, followed by an unit cell of YBCO-124 characterized by double CuO chains between Ba atoms. This alternation of single and double CuO chains effectively forms a region of YBCO-247, which comprises YBCO-123 and YBCO-124 building blocks. These nanoscale YBCO-247 regions appear in patchy strips throughout the YBCO layer, as indicated by the arrows in Fig. 1(b). Further on the right of Fig. 1(c) is another YBCO-247 unit cell, also containing a double CuO chain. It is interesting to note the variation in the Cu atom alignment in the double CuO chains. In the YBCO-247 unit cell on the left, the Cu atoms are horizontally aligned along the $c$-axis; whereas, in the YBCO-247 unit cell on the right, the Cu atoms are staggered in a zigzag fashion. This difference can be interpreted in terms of micro-twinning between the two YBCO-247 unit cells, whose $a$- and $b$-axis orientations are switched. Finally, because the YBCO target used has 123 stoichiometry, the occurrence of double CuO chains is expectedly compensated by missing CuO chains elsewhere in the film. Such a chain-less region can be seen between the two YBCO-247 unit cells, showing two adjacent Ba layers with no CuO chain in between. It should be remarked that such double-chain and missing-chain intergrowths are largely absent in unilayer YBCO films grown under similar conditions, as shown by the STEM image in Figure 2.

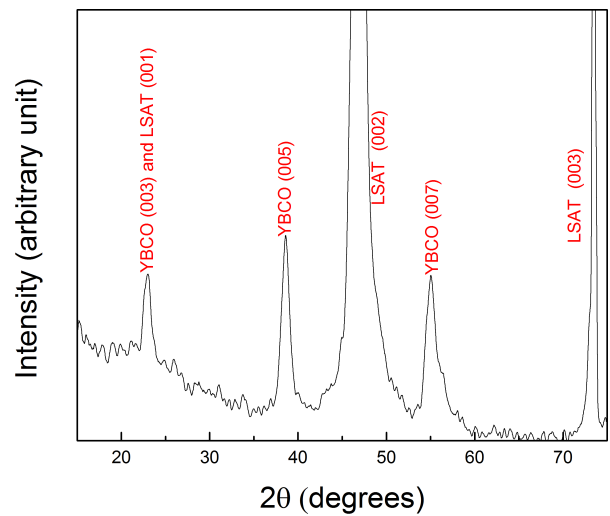

FIG. 3. XRD pattern of a 25 nm/50nm bilayer LCMO/YBCO film grown on (001)-oriented LSAT substrate. Only peaks associated with the c-axis of either YBCO-123, LCMO, or LSAT are visible. No peaks associated with YBCO-247 are visible, indicating that the nanoscale YBCO-247 regions seen in the high-resolution STEM image do not appear in the XRD.

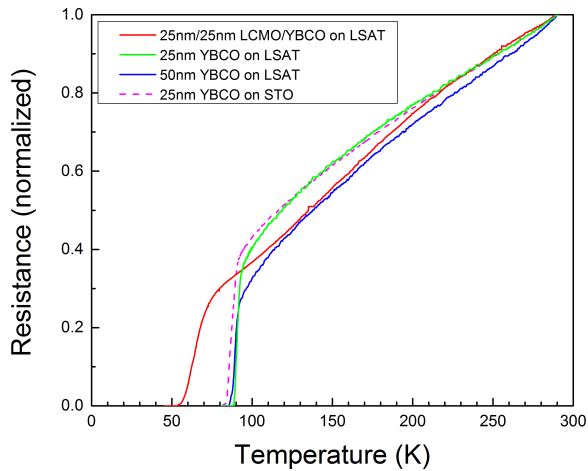

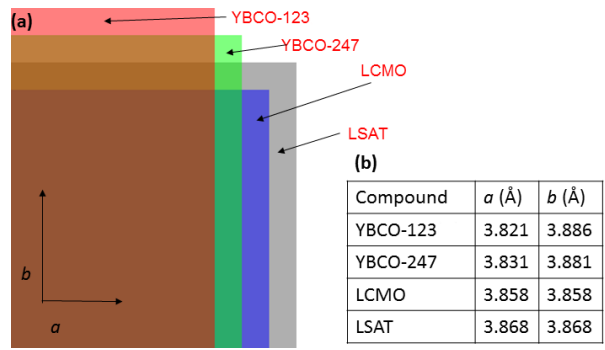

| Compound | $a$ (Å) | $b$ (Å) |
|----------|---------|---------|
| YBCO-123 | 3.821 | 3.886 |
| YBCO-247 | 3.831 | 3.881 |
| LCMO | 3.858 | 3.858 |
| LSAT | 3.868 | 3.868 |

FIG. 5. Comparison of the ab-plane lattice structure between YBCO-123, YBCO-247, LCMO, and LSAT; (a) shows a schematic diagram illustrating the differences in both lattice symmetry and lattice parameters, which are tabulated in (b). In the diagram, the relative length scales between the a- and b-axes for each material, and between the materials, are exaggerated for clarity.

FIG. 4. Plot of normalized resistance vs. temperature for unilayer YBCO and bilayer LCMO/YBCO films. All the unilayer films show sharp superconducting transitions with $T_c$ near 90 K. The 25 nm/25 nm bilayer LCMO/YBCO film shows a significantly reduced $T_c$ near 60 K, with a broadened transition.

To probe the pervasiveness of these nanoscale YBCO-247 regions, we carried out XRD on our LCMO/YBCO thin films. Figure 3 shows the diffraction pattern of a 25 nm/50 nm bilayer LCMO/YBCO film grown on LSAT substrate. All the major peaks are identified in terms of the $c$-axis lattice of either YBCO-123, LCMO, or LSAT, although the peaks for LCMO and LSAT are not distinguishable because of their close lattice parameters. The YBCO (003)-peak at $2\theta = 22.85°$ and the LSAT (001)-peak at $2\theta = 22.99°$ are also not distinguishable. By relating the YBCO (005)-and (007)-peaks with $2\theta = 38.55°$ and $2\theta = 55.06°$, respectively, we find the $c$-axis lattice parameter of our YBCO film to be 11.68Å, in agreement with values reported in the literature.[35] It should be emphasized that no peaks associated with YBCO-247 are visible in the XRD pattern within the resolution of our instrument, indicating that the nanoscale YBCO-247 regions seen in the high-resolution STEM image do not appear in the XRD.

The occurrence of double CuO-chain intergrowths in our bilayer LCMO/YBCO thin films can be physically linked to their resistively measured $T_c$. Figure 4 plots the resistance $R$ vs. temperature $T$ data taken on various films. To facilitate comparison, each $R$ vs. $T$ curve is normalized to its $R$ value at room temperature. Both unilayer YBCO films grown on LSAT show sharp superconducting transitions with $T_c$ near 90 K, consistent with the YBCO being fully oxygenated. As a control, we also grew unilayer YBCO films on $SrTiO_3$ (STO), which show a similar $T_c$ also near 90 K. These results indicate that the resistive $T_c$ of the YBCO layer is largely insensitive to the lattice mismatch with the substrate material, down to 25 nm YBCO thickness. However, the 25 nm/25 nm bilayer LCMO/YBCO film shows a much lower $T_c$, near 60 K, and a broader transition than any of the unilayer YBCO films, indicating that the addition of an epitaxial LCMO overlayer significantly reduces the resistive $T_c$ in the YBCO layer. We can plausibly attribute this $T_c$ reduction to the nanoscale YBCO-247 regions seen in our high-resolution STEM images, since YBCO-247 has generally shown lower $T_c$ than either YBCO-124 or fully-oxygenated YBCO-123.[36,37] We note that an alternative explanation of this $T_c$ reduction in terms of under-oxygenated YBCO-123 is not likely, since the LCMO overlayer was grown *in-situ* at an even higher oxygen pressure than the YBCO layer.[13]

To explain the formation of CuO intergrowths in our LCMO/YBCO films, we consider the lattice structures of the materials involved in the heteroepitaxy. Figure 5 gives a comparison of the *ab*-plane lattice structures between YBCO-123, YBCO-247, LCMO, and LSAT. Fig. 5(a) shows a schematic diagram illustrating the differences in both lattice symmetry and lattice parameters,[38-40] which are tabulated in Fig. 5(b). First, it is well known that all the superconducting phases of Y-Ba-Cu-O are orthorhombic due to the CuO chain that runs along the *b*-axis. Because of the inter-chain attraction within the double-CuO chains, YBCO-247 has a shorter *b*-axes and is thus less orthorhombic than YBCO-123. Since both the LCMO overlayer and the LSAT substrate have cubic lattices, their combined mismatch in lattice symmetry with the YBCO layer would favor the formation of the less orthorhombic YBCO-247 phase. In addition to this bilayer lattice-symmetry mismatch, the *a*- and *b*-axis lattice parameters of both LCMO and LSAT are closer to YBCO-247 than to YBCO-123. Thus the lattice-parameter mismatch, from both sides of the YBCO layer, also tends to favor the formation of YBCO-247. In essence, the intergrowth of double CuO chains provides an effective mechanism for relieving the het-eroepitaxial strain, imposed by both the LCMO overlayer and LSAT substrate, in the YBCO layer.

In summary, we have performed HAADF-STEM, XRD, and electrical resistance measurements on bilayer LCMO/YBCO and unilayer YBCO thin films grown by PLD. The STEM images on the bilayer films revealed YBCO-247 regions formed by double CuO-chain intergrowths, which we attribute to heteroepitaxial lattice mismatch of the YBCO layer with both the LCMO overlayer and LSAT substrate. These nanoscale 247 regions do not appear in XRD, but can physically explain why $T_c$ in the bilayer LCMO/YBCO thin films is significantly lower than 90 K, at least down to $\sim 25$ nm YBCO thickness. Our results suggest an alternative framework, in terms of nanoscale phase inhomogeneity induced by heteroepitaxial lattice mismatch, for understanding the dependence of $T_c$ on layer thicknesses in LCMO/YBCO multilayers. As a microstructural mechanism for the attenuation of superconductivity in LCMO/YBCO heterostructures, the epitaxially-induced CuO intergrowths are also potentially useful as controlled inclusions for patterning cuprate nanostructures.

## ACKNOWLEDGMENTS

This work is supported by NSERC (through Discovery Grants to GAB and JYTW) and CIFAR. The electron microscopy work was carried out at the Canadian Centre for Electron Microscopy, a National Facility supported by McMaster University and NSERC.